\documentclass[10pt,preprint]{aastex}
\usepackage{amsmath}
\usepackage{graphicx}
\usepackage{graphics}
\usepackage[dvips]{color}
\newcommand{\kms}{{km~s$^{-1}$}}

\begin{document}

\title{SDSS J153636.22+044127.0 and its analogues: shocked outflows, not active binary black holes}
\author{Shaohua Zhang\altaffilmark{1}, Hongyan Zhou\altaffilmark{1,2}, Xiheng Shi\altaffilmark{1}, Tuo Ji\altaffilmark{1}, Peng Jiang\altaffilmark{1}, Xiang Pan\altaffilmark{1}, Zhenfeng Sheng\altaffilmark{2}, Luming Sun\altaffilmark{2} and Zhihao Zhong\altaffilmark{1,2}}
\affil{$^1$SOA Key Laboratory for Polar Science, Polar Research Institute of China, 451 Jinqiao Road, Shanghai, 200136, China; zhangshaohua@pric.org.cn, zhouhongyan@pric.org.cn\\
$^2$CAS Key Laboratory for Research in Galaxies and Cosmology, Department of Astronomy,
    University of Sciences and Technology of China, Hefei, Anhui 230026, China}

\begin{abstract}
The binary emission-line system, particularly the binary broad-line emission system, is considered the most effective indicator of the black hole binary. A plausible example of such a system, SDSS J153636.22+044127.0, was reported as the first known object with two hydrogen Balmer broad-line systems, which are interpreted to be the result of broad-line regions around a pair of black holes (Boroson \& Lauer 2009). Here, we show the follow-up optical and near-infrared spectral observations of SDSS J153636.22+044127.0 and its analogues. In these objects, the broad hydrogen Balmer and Paschen, He I and Mg II lines share the same peculiar emission-line profile (including a blue system, a red system and a double-peaked disk-line component); however, the invariance in the large time interval, the absence of the blue system in He I $\lambda$10830 profile and the abnormally strong emission of the hydrogen Pa$\beta$ blue system oppose the binary proposal. We suggest that these unique broad lines arise from the AGN emission-line region and the shock-heated outflowing gases rather than a binary system of two active black holes.
\end{abstract}

\maketitle

\section{Introduction}
The quasar SDSS J153636.33+044127.0 (hereafter SDSS J1536+0441 for short) is reported with great expectation as a candidate for hosting a sub-parsec binary supermassive black hole (SMBH) system (Boroson \& Lauer 2009). This expectation comes from the fact that it is the first known case whose peculiar hydrogen Balmer  profiles exhibit two broad-line emission systems separated by $\rm \sim 3500~km~s^{-1}$ (referred to here as the `blue-system' and `red-system') but only one narrow-line emission system associated with the red-system.  The multiple components were expected to have arisen from two broad-line emission regions (BLRs) around a pair of black holes with masses of $10^{7.3}$  and $10^{8.9}$ M$_{\odot}$ and with a rotation period of $\sim 100$ years.

After the publication of this work, follow-up imaging and spectroscopic observations gradually deepened doubts about the binary interpretation. The images by the VLA  at 8.5 GHz (Wrobel  \& Laor  2009) and by the European VLBI Network at 5 GHz (Bondi \& P{\'e}rez-Torres  2010) resolved two pc-scale cores separated by $0''.97$ ($5.1$ kpc), leading to the possibility of a celestial body pair, which was also confirmed by the Hubble Space Telescope WFPC2/PC images (Decarli et al.\ 2009) and the ESO/Very Large Telescope high-resolution $VzK$ images (Lauer  \& Boroson \ 2009).
The radio luminosity ($L_R \sim 10^{40}\rm ~ erg~ s^{-1}$) and the flat radio spectrum suggest that both radio nuclei are powered by their own AGNs rather than by a 0.1 pc black hole binary (Bondi \& P{\'e}rez-Torres  2010).
The third velocity components identified in the redwing of the red-system H$\alpha$ and H$\beta$ lines suggest that SDSS J1536+0441 is a double-peaked emission-line quasar (Chornock  et al.\  2009).

The further reconstruction of the H$\alpha$ and H$\beta$ profiles (with a circular Keplerian disk-line component and multi-Gaussian components) suggests that the unusual emission system of SDSS J1536+0441 is both a double-peaked emitter and a binary SMBH system and that the extra fluxes in the blue peaks come from the region around the secondary black hole (Tang \& Grindlay \ 2009).
However, over the almost one year difference between the rest-frame epoch of the KPNO Mayall, Palomar Hale, Keck II and SDSS spectroscopies, the lack of velocity evolution and amplitude variation in either broad-line system appears to be too unique among either class of double-peaked emitters and black hole binaries (Lauer \& Boroson 2009; Chornock et al.\  2010).
The questions of whether SDSS J1536+0441 is a binary and of the origin of the unique emission lines, particularly the blue-system, are still unanswered.
From May 2015 to March 2017, we accumulated optical and near-infrared (NIR) spectra of SDSS J1536+0441 on the Palomar 200-inch Hale telescope, which are likely able to unravel the mystery when analyzed with the archives obtained ten years ago.

\section{Observations and Analysis}
The archival optical spectra of SDSS J1536+0441 used in this work were obtained from the  Sloan Digital Sky Survey (SDSS; York et al. 2000) spectroscopic database and the Keck Echellette Spectrograph and Imager (ESI; Sheinis et al. 2002) archive observed by Chornock et al. (2010).
The SDSS spectrum was obtained by the SDSS 2.5-m telescope with an exposure time of 4806 seconds on March 07$\rm ^{th}$, 2008. The 1-D spectrum was accessed from the SDSS Data Release Seven (DR7; Abazajian et al. 2009).
The SDSS spectrographs provide spectra with a high signal-to-noise ratio (SNR) at a resolution of $R \sim 1800$ and a wavelength range from 3800 to 9200 \AA\ (Stoughton et al. 2002).
The Keck ESI spectrum was obtained from a pair of 600-second observations on the Keck II 10-m telescope on March 22$\rm ^{nd}$, 2009. The 0.75 arcsec slit was used, giving a resolution of $\rm \sim45\ km\ s^{-1}$ over the range of $\rm 4000-10200$ \AA.

Our follow-up optical spectrum of SDSS J1536+0441 was obtained with the Double Spectrograph (DBSP; Oke \& Gunn 1982) spectrograph on the 200-inch Hale telescope at Palomar Observatory on March 19$^{\rm th}$, 2017. Two exposures of 400 seconds each were performed, and a 1.5 arcsec slit was used during the night. We chose a 600/4000 grating for the blue side and a $316/7500$ grating for the red side, and a D55 dichroic was selected. The setting yielded a wavelength coverage of $\rm 2778-5818$ \AA\ and $\rm 5358-11652$ \AA\ with median resolutions of $\sim1200$ for the blue side and $\sim 2800$ for the red side. The follow-up optical data were calibrated using the hot subdwarf star Feige 66. The DBSP spectroscopic data were reduced following the IRAF\footnote{IRAF is distributed by the National Optical Astronomy Observatories, which are operated by the Association of Universities for Research in Astronomy, Inc. under a cooperative agreement with the National Science Foundation.} standard routine.
Wavelength calibration was carried out using Fe-Ar and He-Ne-Ar lamps on the same night during the observations. The standard stars were observed for flux calibrations before or after the observations of SDSS J1536+0441.

The NIR spectra of SDSS J1536+0441 were performed with the Triple Spectrograph (TSpec; Wilson et al. 2004) spectrograph on the Palomar 200-inch Hale telescope on May 24$\rm ^{th}$, 2015, and March 13$\rm ^{th}$, 2017. For each observation, four exposures of 300 seconds each were taken in an A-B-B-A dithering model. The TSpec spectrograph provides simultaneous wavelength coverage from 0.9 to 2.46 $\mu$m at a resolution of $\rm 1.4 - 2.9\ \AA$, with two gaps at approximately 1.35 and 1.85 $\mu$m owing to the telluric absorption bands.
Slits with a width of 1.0 arcsec were used, and the spectral resolution is $\sim2200$ in the $H$-band, where the He I $\lambda10830$ emission line is located. The follow-up NIR data were calibrated using the A0V star HD139031. The raw data of two observations were processed using the Spextool software package (Vacca et al. 2003; Cushing et al. 2004).

\begin{figure*}[tbp]
\epsscale{0.6} \plotone{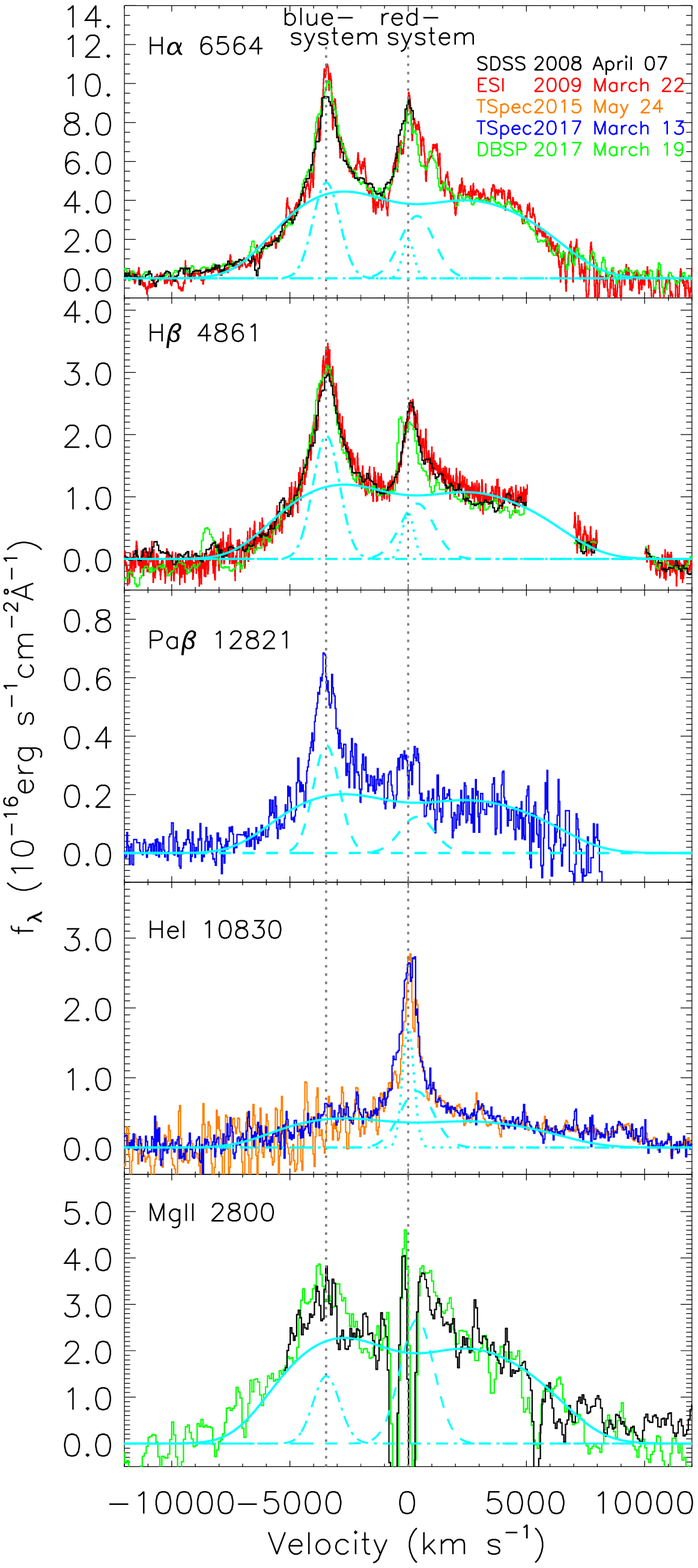}
\caption{Emission-line profiles of the hydrogen Balmer and Paschen, He I $\lambda10830$ and Mg II lines of SDSS J1536+0441. The solid lines are the continuum-removed spectra, which are from the SDSS and Keck archives and the Palomar DBSP and TSpec follow-up observations. In the panels, the solid cyan line is the best-fit disk-line components, the dash-dot and dashed cyan lines are the Gaussian fit of the broad blue-system and red-system, and the dotted line is the Gaussian fit of the narrow lines. The vertical gray dotted lines are the centers of the blue-system and red-system.}\label{fig-J1536lines}
\end{figure*}

In order to study the properties of the emission line, a simple spline fit to the continuum beneath each emission line was first subtracted following the method  of Chornock et al. (2010).
The continuum windows of [2580, 2670] \AA\ and [2890, 2950] \AA\ for Mg II, [4320, 4720] \AA\ and [5080, 5300] \AA\ for H$\beta$, [6150, 6250] \AA\ and [6850, 6950] \AA\ for H$\alpha$,  [1.03, 1.06] $\mu$m and [1.10, 1.12] $\mu$m for He I $\lambda10830$,
and [1.19, 1.24] $\mu$m  for Pa$\beta$ were chosen.
After that, in Figure \ref{fig-J1536lines}, the multi-epoch emission-line profiles of H$\alpha$, H$\beta$, Pa$\beta$, He I $\lambda10830$, and Mg II are displayed in the velocity space at $z = 0.3889$. The blue-system and red-system are also marked by the vertical dotted lines.
Two early observations of this are selected from the SDSS and Keck ESI archives.
The addition of the newest spectrum from the DBSP offers us an opportunity to effectively check the profile change in a long temporal baseline ($\Delta T\rm \simeq  9~yrs$), since the binary system would have a velocity shift in the blue-system of $\rm \sim 100\ km\ s^{-1}$ in a single year based on the projection model (Boroson \& Lauer 2009). Although the original SDSS spectrum did not cover the full H$\alpha$ and Mg II profiles, comparisons are sufficient to present an unmistakable conclusion: there is no visible velocity shift or amplitude change in the blue-system and red-system, even though they were found in previous works (Lauer \& Boroson 2009; Chornock et al. 2010).
This is obviously inconsistent with the Boroson and Lauer's prediction.
However, intriguingly, two infrared spectroscopic observations show that the He I $\lambda10830$ emission line seems to have a similar disk-line component to other lines and only one emission-line peak with the same velocity as the red-system (at $\rm \sim 0\ km\ s^{-1}$); on the contrary, the emission-line profile of hydrogen Pa$\beta$ exhibits (almost 3 times) stronger emission of the blue-system than of the red-system. The deviant flux ratios of the blueshifted emission-line component would have a serious effect on the binary hypothesis in the following analysis.

As described in Tang and Grindlay (2009), the complex emission-line profile is decomposed to a bottom double-peaked emitter (e.g., Eracleous et al. 1997) plus two extra emission-line systems. The former is the disk-line component from the accretion disk around the primary black hole, and the latter was interpreted as a binary black hole system (Boroson \& Lauer 2009). 
In this work, we follow the method of Tang and Grindlay (2009) to describe the emission-line profile: an axisymmetric Keplerian disk-line model from Chen and Halpern (1989) for the bottom double-peaked component, one Gaussian for the blue-system and two (broad and narrow) Gaussians for the red-system.
With the exception of the flux strengths, other parameters for each emission line, i.e., five disk-line model parameters, velocity shift and the width of each system, are taken directly from Tang and Grindlay's best-fit for H$\alpha$\footnote{In this work, the emission-line fluxes are estimated under the assumption that the emission-line component emitting from the same gas cloud will have the same profile. We also noticed that Tang \& Grindlay (2009) showed the different widths for the disk-line bottom and the red-system in H$\alpha$ and H$\beta$ lines (H$\beta$ is suggested broader than H$\alpha$). However the larger width of the red-system in H$\beta$  is probably due to the underestimation of disk-line bottom in H$\beta$ at around 0 \kms. In fact, it is hard to really distinguish the disk-line models of H$\alpha$ and H$\beta$ with current data. If the best-fit disk model for H$\alpha$ is used in H$\beta$ profile fitting, the width of the red-system  in H$\beta$ could be almost the same as that in H$\alpha$. }.
In particular, the disk inclination is $i \sim 47^\circ$, the inner/outer radii are $r_1 \sim 1000\ r_G$ and $r_2 \sim 13,000\ r_G$, the velocity dispersion is $\rm \sigma=1200\ km\ s^{-1}$, and the surface emissivity power law is $q=-3$. Furthermore, the velocity shifts of the blue-system and the broad and narrow components of the red-system are $-3470$, 420 and $\rm 0\ km\ s^{-1}$, and their $FWHM$ values are 1340, 1690 and $\rm 573\ km\ s^{-1}$, respectively. More details can be seen in Section 2 and Table 1 of Tang and Grindlay (2009). This set of parameters also applies to the measurements of H$\beta$, Mg II and He I $\lambda10830$. The fitting results are overplotted by cyan in Figure \ref{fig-J1536lines} for each line with least $\chi^2$, and the emission fluxes for each component are listed in Table \ref{tab1}.

\begin{figure*}[tbp]
\epsscale{1.0} \plotone{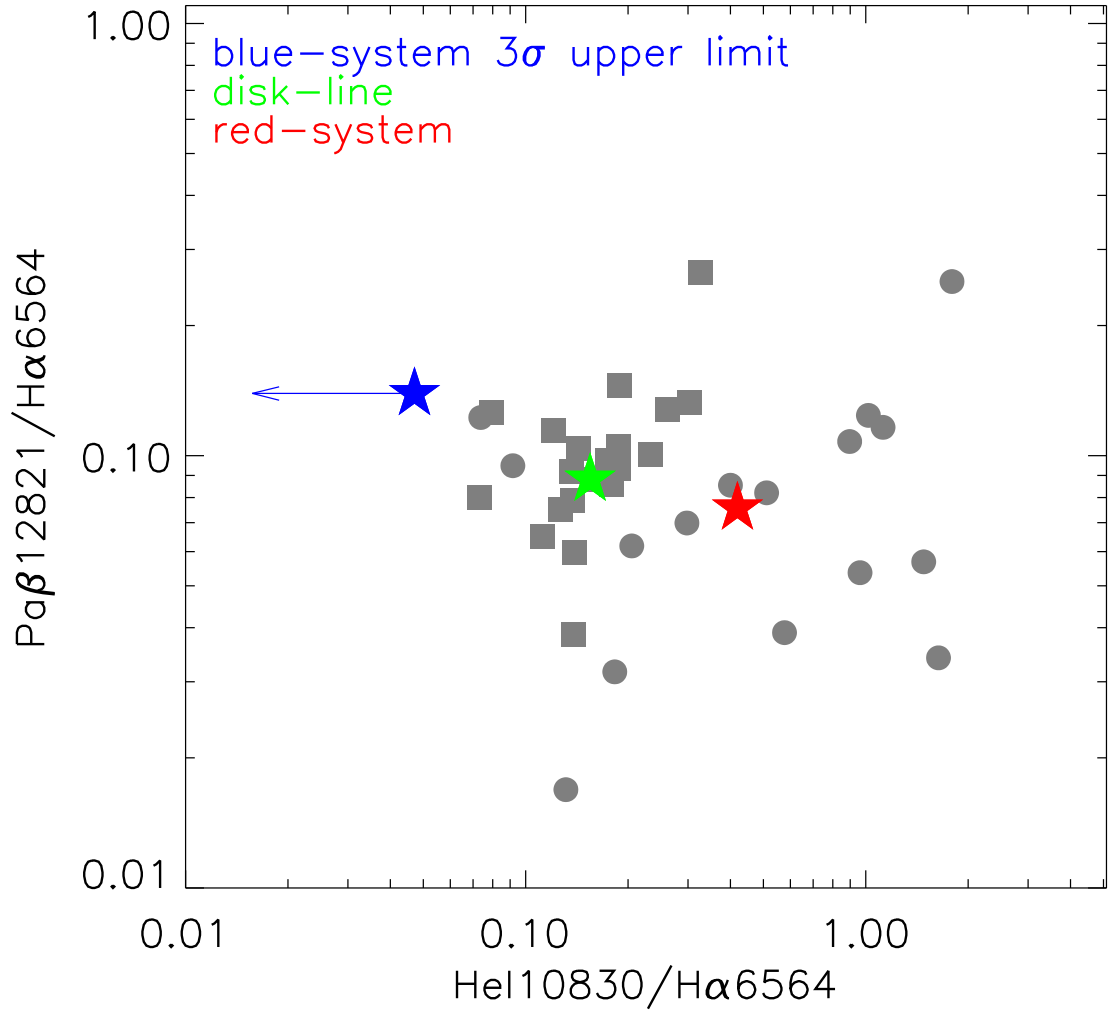}
\caption{Observed flux ratio comparison of the He I $\lambda$10830, Pa$\beta$  and H$\alpha$ lines of SDSS J1536+0441 with the 21 well-known broad-line AGNs presented in Landt et al. (2008). The green, blue and red stars represent the disk-line component, blue-system and red-system of SDSS J1536+0441, respectively. The gray squares and circles show the flux ratios of the broad and narrow emission lines of the comparison sample. The blue-system of SDSS J1536+0441 is obviously an outlier in the two-dimensional chromatogram of emission-line flux ratios.}\label{fig-lineratio}
\end{figure*}

In Figure \ref{fig-lineratio}, we present the flux ratios of He I $\lambda10830$, Pa$\beta$ and H$\alpha$ lines of the disk-line component, blue-system and red-system.  Since the He I $\lambda10830$ line is not visible to the naked eye in the blue-system, its $3\sigma$ upper limit is estimated to be $\rm 7.4 \times 10^{-17}~erg~s^{-1}~cm^{-2}$ from the H$\alpha$ emission-line width and the observed flux errors\footnote{For a Gaussian-like emission-line profile, the total flux roughly approximates $f_{\rm max} \times FWHM$, where $f_{\rm max}$ and $FWHM$ are the peak flux density and the full width at half maximum of the emission-line profile.
If we suppose the the 1-$\sigma$ fluctuation of the observed $f_{\lambda}$, $\sigma_f$, to be the 1-$\sigma$ upper limit for the peak flux density of undeteced He I $\lambda$10830, the 1-$\sigma$ upper limit for the total flux could then be defined as $\sigma_{f}\times FWHM \approx \int_{FWHM} \sigma_{f} {\rm d}\lambda$.}.   Landt et al. (2008) measured and presented the permitted transitions of hydrogen, helium, oxygen, and calcium and the rich spectrum of singly ionized iron for 23 well-known broad-line AGNs in their high-quality optical and NIR spectra (Table 5 in their work); the flux ratios of broad/narrow He I $\lambda10830$, Pa$\beta$ and H$\alpha$ lines of 21 of these are also displayed for comparison. The blue-system is obviously an outlier in the two-dimensional chromatogram of emission-line flux ratios.
This is not a unique instance; other similar systems have been reported. Two other similar cases---SDSS J132052.19+574737.3 and SDSS J150718.10+312942.5---are found in the SDSS archive, the emission lines of which show almost the same emission-line profiles as SDSS J1536+0441. It is inconceivable that the H$\alpha$ emission line contains the disk-line component,  blue-system and red-system  but that the blue-system of the He I $\lambda10830$ profile is absent.

\section{Discussion}

\begin{figure*}[tbp]
\epsscale{1.0} \plotone{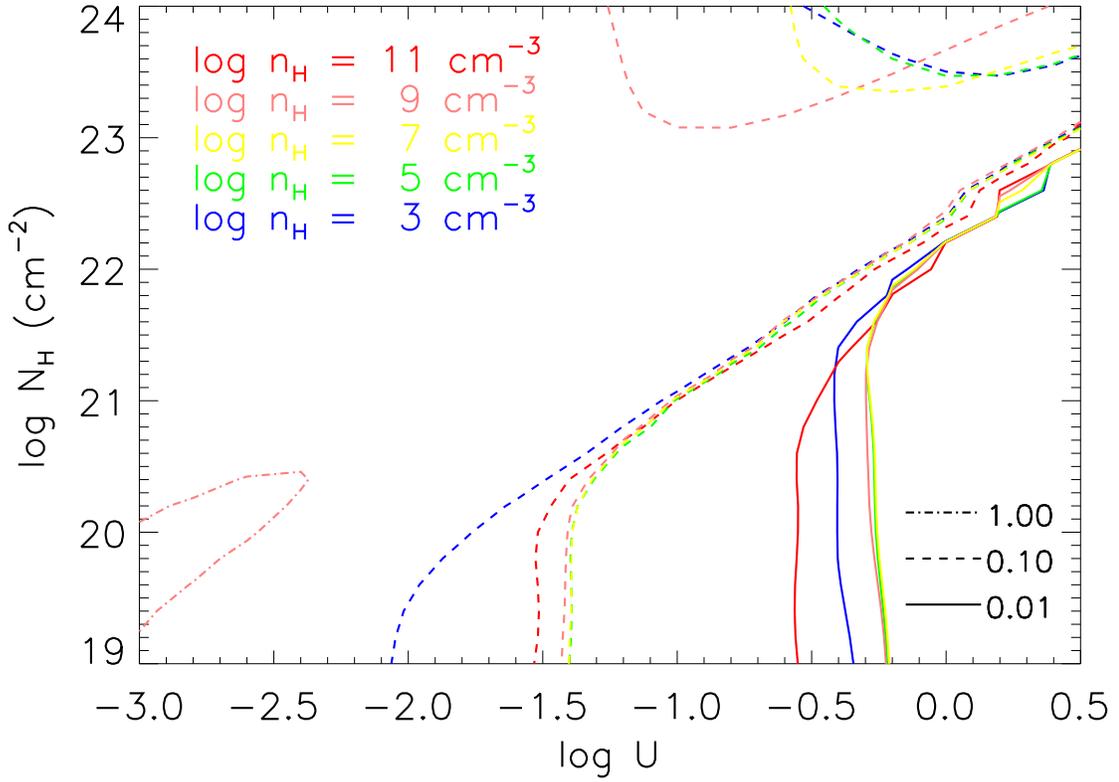}
\caption{Photoionization models of the flux ratio (HeI10830/H$\alpha$6546) irradiated by the AGN's central  ionization continuum. The flux ratios with different colors are presented in the log $U{\rm ~(ionization~ parameter)~-~log}~N\rm{_{H}~(column~ density)}$ space for the five typical gas densities (${\rm log}~n{\rm_{H}~(cm^{-3})}=$ 3, 5, 7, 9, and 11). The extensive parameter space is almost  sufficient to cover all the routine possibilities of the AGN `normal' emission-line regions.
However, the typical BLR gases with $n\rm_{H}\sim10^9-10^{10}~cm^{-3}$,  $N\rm_{H}\sim10^{22}~cm^{-2}$, and $U\rm\sim10^{-2}-10^{-1}$ present relatively strong He I $\lambda10830$ emission. The value of $\rm HeI10830/H\alpha6564$ is at least larger than 0.1, which is two times the observed $3\sigma$ upper limit.}\label{fig-AGNmodel}
\end{figure*}

To investigate the emission-line properties in the blue-system, we employed the photoionization code $Cloudy$ (Version 17.01, Ferland et al. 2017) and applied the measured emission-line ratios to these models to simulate the possible physical conditions and processes in the medium. The simple model is a slab-shaped gas with a unique density and homogeneous chemical composition of solar values, irradiated directly by the central ionization continuum source.
For the broad-line region (BLR) around one of the hypothesized binary black holes, such primordial models can be simply described using parameters such as density $n_{\rm H}$, hydrogen column density $N_{\rm H}$, ionization parameter $U$, and spectral energy distribution (SED) of the incident radiation. A typical AGN ionization continuum is applied as the incident SED, which is a combination of a blackbody ``Big Bump'' and power laws\footnote{See details in Hazy, a brief introduction to Cloudy; http://www.nublado.org.}.
The component peaks  at $\approx$ 1 Ryd and is parameterized by $T\rm_{BB}=1.5\times10^5$ K. The slope of the X-ray component, the X-ray to UV ratio, and the low-energy slope are set as $\alpha_{\rm x}=-2$ (beyond 100 keV) and $-1$ (between 1.36 eV and 100 keV),  $\alpha_{\rm ox}=-1.4$, and $\alpha_{\rm UV}=-0.5$, respectively. This UV-soft SED is regarded as more realistic for radio-quiet quasars than the other available SEDs provided by $Cloudy$  (see the detailed discussion in Section 4.2 of Dunn et al. 2010). We calculated a series of photoionization models with different ionization parameters, densities and hydrogen column densities. The ranges of these parameters are $ -4.0 \leq {\rm log_{10}}~U \leq 1.0$, $3.0 \leq  {\rm log_{10}}~n_{\rm H}~({\rm cm^{-3}})  \leq  11.0$ and $19.0 \leq  {\rm log_{10}}~N_{\rm H}~({\rm cm^{-2}})  \leq  24.0$ with a step of 0.2 dex, which could well cover the possible parameter space of the broad- and narrow-line regions. In Figure \ref{fig-AGNmodel}, the flux ratios of HeI$\lambda$10830/H$\alpha$ are presented in the ${\rm log~}U- {\rm log~} N_{\rm H}$ space for the five densities (${\rm log~n_{\rm H}~(cm^{-3})}=$ 3, 5, 7, 9, and 11).
The extensive parameter space is almost enough to cover all the routine possibilities of the AGN `normal' emission-line regions. However, the typical BLR gases with $n\rm_{H}\sim10^9-10^{10}~cm^{-3}$,  $N\rm_{H}\sim10^{22}~cm^{-2}$ and $U\rm\sim10^{-2}-10^{-1}$ present the relatively strong He I $\lambda10830$ emission. If the blue-system of SDSS J1536+0441 is also compared, the observed ratio of $\rm HeI10830/H\alpha6564$ is at least larger than 0.1, which is two times the $3\sigma$ upper limit. Further extensive calculation in a larger parameter space suggests that a parameter combination range of $U n\rm_{H} \approx 10^{11.5}~cm^{-3}$ with an exceedingly high density of $n_{\mathrm{H}} \ge 10^{12}~\mathrm{cm}^{-3}$ would reproduce the observed flux ratios of the hydrogen and helium lines. If the blue-system emits from such a high-density medium, the size would be 50 times less than that of the red-system BLR (based on the sub-parsec binary hypothesis, in which both BLRs are illuminated by the same ionizing flux), and the number ratios of ions in the blue-system and red-system  would be only approximately 5 per thousand. Even accounting for the higher emission efficiency (within an order of magnitude), the high-density medium is not sufficient to emit blue-system lines comparable to those of the red-system.

In previous works, we accurately measured the blueshifted emission-line system in two quasars---SDSS J000610.67+121501.2 (Zhang et al. 2017) and SDSS J163459.82+204936.0 (Liu et al. 2016)---and those of IRAS 13224-3809 and 1H 0707-495 were reported by Leighly et al. (2004). In general, the blueshifted lines in these cases are shifted approximately one to two thousand kilometers per second, but their emitting winds are also irradiated by the AGN ionization continuum; therefore, the blue-system of SDSS J1536+0441 does meet the scenario of outflowing emission gases. However, the interaction of the outflowing winds with the peripheral expanding matter provides other possible ionizing sources. A probable candidate is the photoionizing shock originating in the collision between massive outflow and the inner surface of the dusty torus. The ionizing photons are produced in the postshock plasma when it cools and diffuse to ionize adjacent medium. In this way, the kinetic energy would eventually be dissipated through radiation. Since the outflow can be accelerated to more than several thousand kilometers per second, the shock velocity $v_{\rm s}$ must be very large. In such a fast shock, the ionizing flux diffusing upsteam forms a photoionization front with a velocity exceeding $v_{\rm s}$. The photoionization front would thus be driven into the preshock gas, expanding as a precursor H II region ahead of the shock. Additionally, emission lines from the precursor H II region could dominate the optical emission of the shock.

The theoretical study of photoionizing shock began with the cloud-cloud collision in BELR and later covered expansion of H II regions, stellar and AGN outflows, supernova blasts, and the collision of galaxies. Great efforts have been performed to develop code simulating these process (Sutherland et al. 1993; Allen et al. 2008). The ionizing continuum itself and the emission lines in both the shock and precursor could be evaluated. Due to the nature of the problems being studied previously, the simulation mainly covers the case of low preshock density ($n{\rm_{H} \le 10^3~ cm}^{-3}$) and low shock velocity ($v_{\rm s}\le 1000~{\rm km~s}^{-1}$). In the low-density shock, various optical forbidden lines, such as [Fe II] $\lambda\lambda$9573,9812,10132, $\lambda\lambda$9956,9998, $\lambda\lambda$10491,10502,10863,11126, and $\lambda\lambda$12570,12791, are extensively used for the diagnosis of shocks. However, all these lines are absent in the blueshifted emission component in SDSS J1536+0441. Therefore, we hypothesize that the preshock density could be much higher, a component that the previous theoretical works did not explore.

Alternatively, we use a toy model to estimate the properties of the shock in SDSS J1536+0441. Since the majority of continuum radiation produced in the cooling zone of photoionizing shock is thermal bremsstrahlung radiation, we use the continuum radiation from optically thin corona as the ionizing spectra, which is also dominated by thermal bremsstrahlung radiation and characterized by the temperature $T$. The SED of incident radiation in our model is generally consistent with the ionizing spectra evaluated by Sutherland et al. (1993). Unlike the typical AGN continuum, for high temperatures of $10^6 < T < 10^8~\mathrm{K}$, the ionizing spectrum peaks at EUV to the soft X-ray band (100 to 1~\AA) and falls rapidly at higher or lower energies. The photoionized medium (precursor H II region) is also described using the three parameters of the ionized gases ($U$, $n_{\rm H}$, and $N_{\rm H}$). Furthermore, the intrinsic value of $\rm H\alpha/H\beta$ was adopted to be 3.1 for active galaxies and 2.85 for H II region galaxies (Veilleux et al. 1987); the observed large flux ratio ($\rm H\alpha/H\beta=3.35\pm0.41$ for the blue-system, and $4.85\pm0.83$ for the red-system) suggests that both emitting gases are probably obscured by the dusty torus. Therefore, an extra parameter, the external dust extinction $E(B-V)$ under the SMC extinction law, is added to the simulation process.

 As we expect in a quite dense medium, $n_{\rm H}$ is set to be larger than $10^6~\rm cm^{-3}$, while $-4\le \log~ U\le 4$, $19 \le \log~ N\rm_{H}~(cm^{-2}) \le 24$, and $6\le \log~ T(\rm K)\le 8$ for the ionizing radiation.
Two unique and clean emission lines (He I $\lambda10830$ and  Pa$\beta$) and the strongest emission line (H$\alpha$) are chosen to investigate the properties of the gas. Their flux ratios, i.e., Pa$\beta$12821/H$\alpha$6564 and HeI$\lambda10830$/H$\alpha$6564 (upper limit), present critical criteria for the parameters, especially $n_{\mathrm{H}}$ and $T$. Theoretical calculations suggest that only in the cases with high-density ($n_{\mathrm{H}} \ge 10^{12}~\mathrm{cm}^{-3}$) gas illuminated by a high-temperature ($T \ge 10^{7}~\mathrm{K}$) shock could the observed values for the line ratios be reproduced.
The observed flux ratios of Pa$\beta$, He I $\lambda10830$ and H$\alpha$ are presented with $n\rm_{H}=10^{12}~cm^{-3}$, $T\rm=10^7~K$, and $E(B-V)=0.3$, 0.4, 0.5 and 0.6. 
They exhibit different variation properties in the ${\rm log}~U - {\rm log}~ N\rm_{H}$ space with the enhancement of extinction; i.e., the allowable parameter space of Pa$\beta$12821/H$\alpha$6564 moves sharply to the highly ionized region, but that of the HeI$\lambda10830$/H$\alpha$6564 (upper limit) exhibits almost no change. The long and narrow overlapping regions in the panels are the possible parameter space for the blue-system emission lines of SDSS J1536+0441. Furthermore, participation of $\rm H\beta4861/H\alpha6564$ further compresses the abovementioned regions and rules out the probability of the extinction $E(B-V)\ge 0.6$.

For example, the simulation-predicted H$\alpha$ luminosity is $\rm 2.33\times10^{7}~erg~s^{-1}~cm^{-2}$
for $T\rm=10^7~K$, $n_{\rm H}=10^{12}~\rm{cm}^{-3}$, $U = 10^{-1.5}$, $N_{\rm {H}}=10^{21.5}~\rm{cm}^{-2}$, and $E(B-V)=0.4$.
The measured H$\alpha$ luminosity of the blue-system is $L\rm_{H\alpha}= (1.11\pm0.05)\times10^{43}~erg~s^{-1}$, so the area of the precursor ionizing region  is $\rm (4.78\pm0.21) \times 10^{35}~ cm^2$, approximately a square with dimensions of 0.22 pc.
Supposing the nucleus continuum follows the red-system broad-line extinction, the corrected bolometric luminosity, $L\rm_{bol} = 2.14\times10^{46}~erg~s^{-1}$,
estimated from the monochromatic luminosity at 5100 \AA\ (Runnoe et al. 2012), suggests that the theoretical sublimation radii of the dust grains would be $R\rm_{in}=1.78~ pc$ ( Kishimoto et al. 2011; Burtscher et al. 2013). If this value is thought of as the inner side of the dusty tours, the precusor ionizing region we observed is only a small part of the surface of the torus. In the general picture of the AGN outflow, the outflowing gases are prevalent in each AGN's structure with a covering factor (Elvis 2000). Thus the collision shock would often be produced on or in the the inner surface of the dusty torus.  However, such cases, i.e, SDSS J1536+0441 and its analogues, are extremely rare, and this phenomenon therefore may require an extreme observation angle to see the emission from the inner surface of the dusty torus through the clearance of the clumpy torus.

\section{Summary}
In this work, SDSS J153636.22+044127.0 and its analogues show that the dual broad emission-line system is likely to be a false-positive indicator of the black hole binary. There are actually many physical mechanisms to produce the blueshifted broad emission-line system, which are more likely to be the result of the AGN-driven or shock-heated outflowing gases. Furthermore, experience has taught us that exploring the black hole binary from the emission-line profiles at optical wavelengths is limiting. For this work, spectral observations are appropriately extended to infrared (and/or ultraviolet) bands, and detection of as many high- and low-ionization emission lines as possible can help us effectively identify the true source of the anomalous emission-line system. Moreover, since possible BLRs' mutual revolution following the orbital motion of the binary system would predict a velocity change in the blue-system, long-time-scale monitoring of the emission-line profile variation can further confirm the truthfulness of the binary black holes.

\acknowledgments
This work is supported by National Natural Science Foundation of China (NSFC-11573024, 11473025, 11421303).
We acknowledge the use of the Hale 200-inch Telescope at Palomar Observatory through the Telescope Access Program (TAP), as well as the archived data from the SDSS and the Keck archives. TAP is funded by the Strategic Priority Research Program, the Emergence of Cosmological Structures (XDB09000000), the National Astronomical Observatories, the Chinese Academy of Sciences, and the Special Fund for Astronomy from the Ministry of Finance. Observations obtained with the Hale Telescope at Palomar Observatory were obtained as part of an agreement between the National Astronomical Observatories, the Chinese Academy of Sciences, and the California Institute of Technology.
Funding for the SDSS and SDSS-II has been provided by the Alfred P. Sloan Foundation, the Participating
Institutions, the NSF, the US Department of Energy, the National Aeronautics and Space Administration (NASA), the Japanese Monbukagakusho, the Max Planck Society, and the Higher Education Funding Council for England. The SDSS website is http://www.sdss.org/.

\clearpage

\figurenum{1}

\figurenum{2}

\figurenum{3}

\figurenum{4}
\begin{figure*}[tbp]
\epsscale{0.9} \plotone{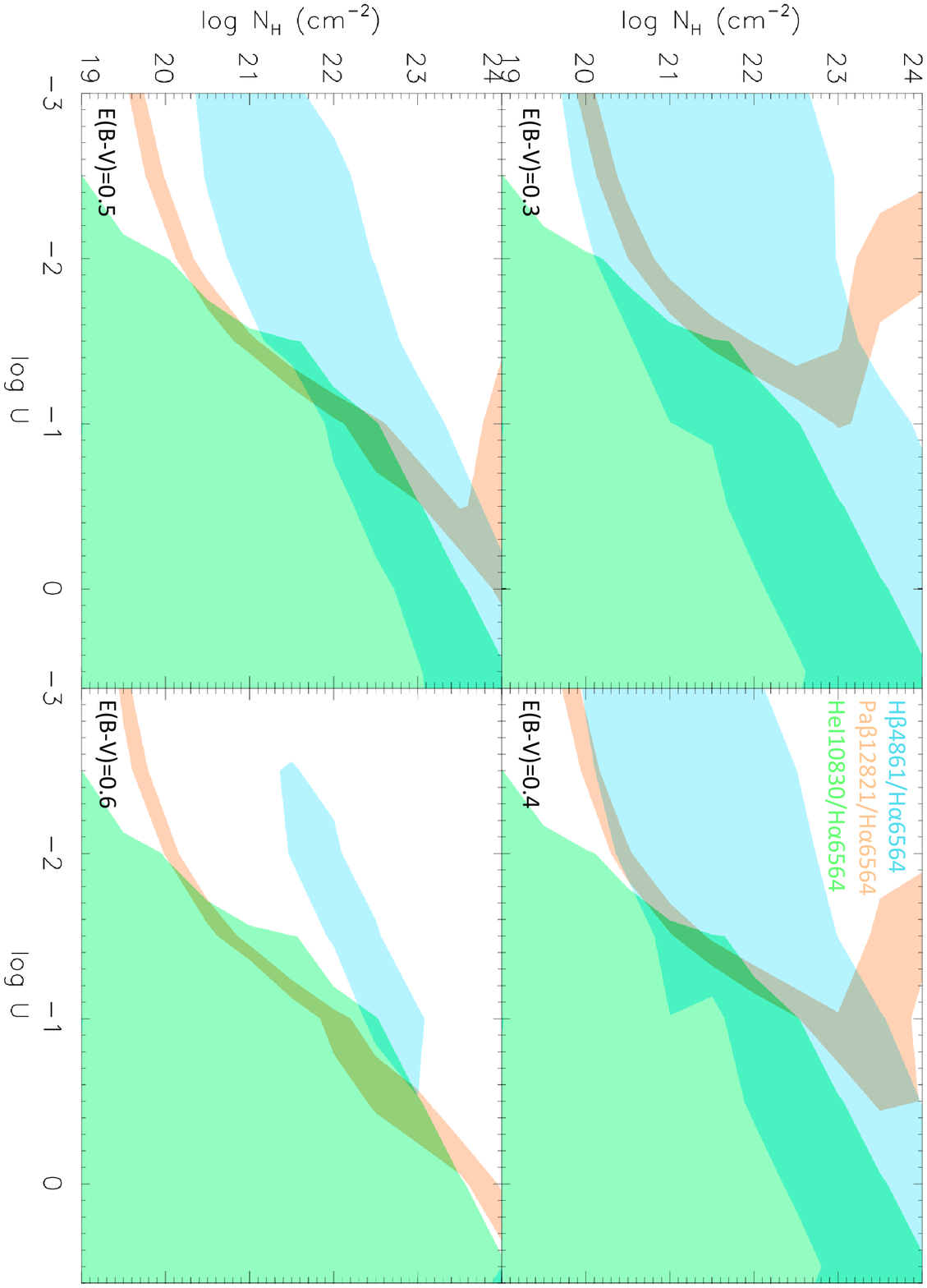}
\caption{Photoionization models of the flux ratios (hydrogen Balmer and Paschen, and He I $\lambda10830$ relative to H$\alpha$) illuminated by the high-temperature ($T = 10^7$ K) shock. The observed 1$\sigma$ uncertainty ranges are presented with density $n\rm_{H}=10^{12}~cm^{-3}$, temperature $T\rm=10^7~K$, and  extinction $E(B-V)=0.3$, 0.4, 0.5 and 0.6. The long and narrow overlapping regions in the panels are the possible parameter space for the blue-system emission lines of SDSS  J1536+0441. The allowable parameter space of Pa$\beta$12821/H$\alpha$6564 moves sharply to the high ionized region, but that of the HeI$\lambda10830$/H$\alpha$6564 (upper limit) exhibits almost no change, and participation of H$\beta$4861/H$\alpha$6564 further compresses the overlapping regions and rules out the excessive extinction.}\label{fig-lineratioshock}
\end{figure*}

\begin{deluxetable}{c c c c c}
\tabletypesize{\scriptsize}
\tablewidth{0pt}
\tablenum{1}
\tablecaption{Emission Line Fluxes of SDSS J153636.22+044127.0
\label{tab1} }
\tablehead{
\colhead{Line} &  \colhead{Disk-line} & \colhead{Blue-system}& \colhead{Red-system} &\colhead{Red-system}\\
\colhead{    } &  \colhead{         } & \colhead{           }& \colhead{broad line} &\colhead{narrow line}\\
\colhead{    } &  \colhead{($\rm 10^{-17}~erg~s^{-1}~cm^{-2}$)} & \colhead{($\rm 10^{-17}~erg~s^{-1}~cm^{-2}$)}& \colhead{($\rm 10^{-17}~erg~s^{-1}~cm^{-2}$)} &\colhead{($\rm 10^{-17}~erg~s^{-1}~cm^{-2}$)}  }
\startdata
H$\alpha$ 6564.02 &  $10873\pm159$  &  $1519\pm68$  &  $1260\pm75$ &  $263\pm43$\\
H$\beta$  4862.62 &  $2161\pm97$    &  $453\pm52$   &  $260\pm42$  &  $73\pm27$\\
Pa$\beta$ 12821.6~~~&  $959\pm84$     &  $220\pm18$   &  $95\pm15$   &  $-$      \\
He I       10833.2~~~&  $1683\pm61$    &  74$^\dagger$ &  $523\pm26$  &  $380\pm23$\\
Mg II 2796,2803&  $2370\pm112$   &  $192\pm34$   &  $401\pm35$  &  $-$
\enddata
\tablenotetext{\dagger} {~the  $3\sigma$ upper limit for the blue-system of He I $\lambda10830$.}
\end{deluxetable}

\clearpage

\appendix
APPENDIX
\section{SDSS J132052.19+574737.3}
\begin{figure}[hbt]
\begin{center}
\includegraphics[width=17cm]{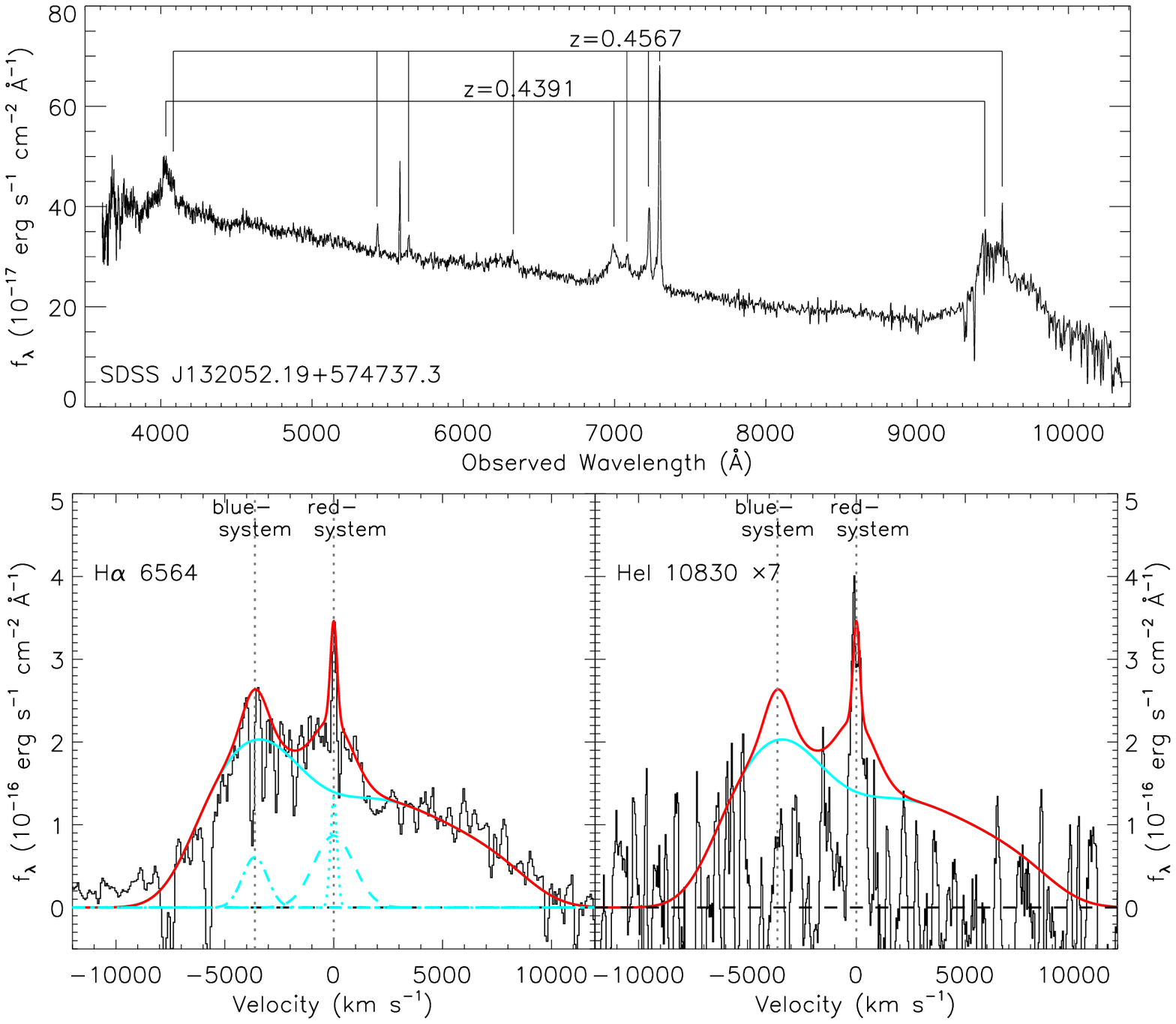}
\label{f1s-J1320}
\end{center}
{\bf{ Figure A1:}} Top: The observed spectrum from the SDSS archive of SDSS J132052.19+574737.3. The two redshifted systems are presented, and the identified features are marked.
The rred-system, at $z=0.4567$, shows typical broad and narrow lines seen in quasars, including
the broad Balmer and Mg II lines and the strong forbidden lines of [O II], [O III], and [Ne III].
The blue-system, at $z=0.4391$, shows only broad Balmer and Mg II  lines.
Bottom: Emission-line profiles of H$\alpha$ and He I $\lambda$10830 (7 times larger) of SDSS J1536+0441. The black lines are the continuum-removed spectra, and the red lines are the sum of the following decomposition components for H$\alpha$.
The solid cyan line is the best-fit disk-line component, the dash-dot and dashed cyan lines are the Gaussian fit to the broad blue-system and red-system, and the dotted line is the Gaussian fit of the narrow H$\alpha$. The vertical gray dotted lines are the centers of the blue-system and red-system.
\end{figure}

As an analogue of SDSS J1536+0441, the SDSS archive provides the optical spectrum of SDSS J132052.19+574737.3 (hereafter SDSS J1320+5747 for short), which was obtained with an exposure time of 4504 seconds on Jun 10$\rm ^{th}$, 2013.
The SDSS archive shows two broad-line emission systems. The higher redshifted `red-system' at $z=0.4567$ shows typical broad (H$\alpha$, H$\beta$, H$\gamma$ and Mg II) and narrow ([O II], [O III] and [Ne III]) lines. The lower redshifted `blue-system' at $z=0.4391$ shows broad Balmer and Mg II lines. The full SDSS spectrum is shown in the top panel of Figure \ref{f1s-J1320}.
To check the emission-line profile of He I $\lambda10830$, we obtained its NIR spectrum with the Palomar TSpec spectrograph on January 21$\rm ^{st}$, 2018. Two runs of four exposures of 120 seconds each in an A-B-B-A dithering model were performed with a slit width of 1.0 arcsec. The follow-up data were calibrated using the A0V star HD116405.
The underlying continua of H$\alpha$, H$\beta$ and He I $\lambda10830$ are subtracted using a simple spline fit to the continuum windows, and the emission-line profiles of SDSS J1320+5747 with the underlying continua removed are shown in the bottom panels of Figure \ref{f1s-J1320}. For a clearer comparison, the emission-line profile of He I $\lambda10830$ is multiplied by 7. H$\beta$ has a similar profile to H$\alpha$, but with a higher emission peak in the blue-system. The same fitting process used by Tang and Grindlay (2009) was used for the emission-line decomposition for H$\alpha$ profile.
For the disk-line model, the velocity dispersion is set to $\rm \sigma=1200\ km\ s^{-1}$, which is a typical value for double-peaked emitters (Strateva et al. 2003), and we fix $q=-3$, as predicted in photonionization calculations (Collin-Souffrin \& Dumont 1989; Dumont \& Collin-Souffrin 1990). The disk inclination is $i \sim 50^\circ$, the inner radius is $r_1 \sim 700\ r_G$, and the outer radius is $r_2 \sim 10,000\ r_G$. When we use the combination of one Gaussian for the blue-system and two (broad and narrow) Gaussians for the red-system to model the extra emission-line components, the velocity shifts of the blue-system and the broad and narrow components of the red-system are $-3620$, 0 and $\rm 0\ km\ s^{-1}$, and their $FWHM$ values are 1345, 2170, and $\rm 354\ km\ s^{-1}$, respectively. In the left panel of Figure \ref{f1s-J1320}, the best-fit components and the sum of the H$\alpha$ emission line are shown in cyan and red. Moreover, we also plot the disk-line component and the modeled profile of H$\alpha$ in the He I $\lambda10830$ panel. Unfortunately, any evidence of broad emission line and disk-line components is undetectable in such a low-quality spectrum, which only catches the narrow He I $\lambda10830$ line.

\section{SDSS J150718.10+312942.5}
\begin{figure}[hbt]
\begin{center}
\includegraphics[width=17cm]{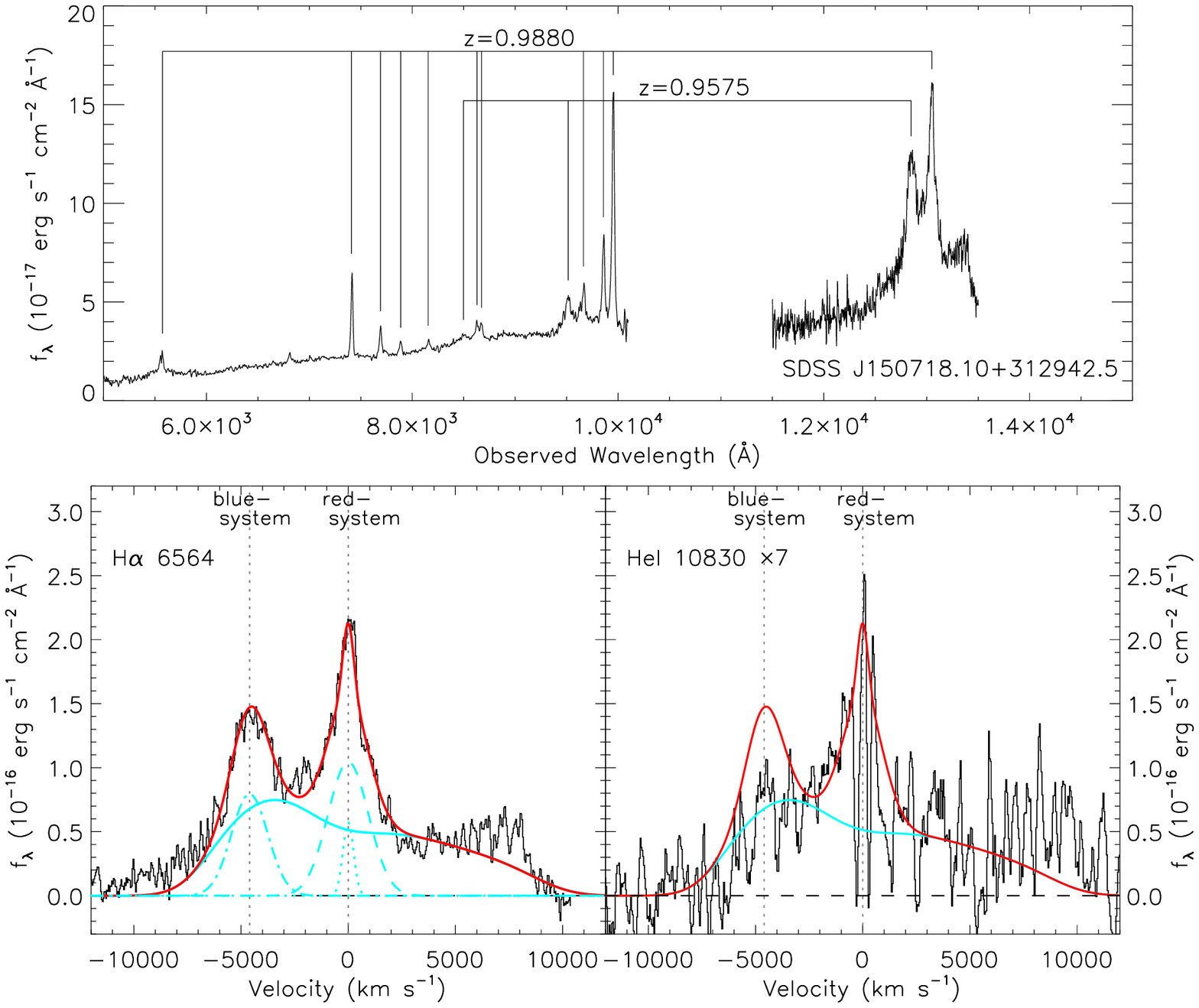}
\label{f1s-J1507}
\end{center}
{\bf{ Figure A2:}} Top: The observed spectrum from the combined archive spectrum and the follow-up high-quality H$\alpha$ emission line of SDSS J150718.10+312942.5. The two redshifted systems are presented, and the identified features are marked. The rred-system, at $z=0.9880$, shows typical broad and narrow lines seen in quasars, including
the broad Balmer and Mg II lines and the strong forbidden lines of [O II], [O III], and [Ne III].
The blue-system, at $z=0.9575$, shows only broad Balmer lines.
Bottom: the same as Figure \ref{f1s-J1320}.
\end{figure}

Optical and NIR spectroscopic observations of SDSS J150718.10+312942.5 (hereafter SDSS J1507+3129 for short and also named as F2M1507+3129), one of 17 highly reddened FIRST-2MASS quasars (Glikman et al. 2004), were carried out on the Keck II 10-m telescope by using the ESI and at the NASA Infrared Telescope Facility (IRTF) by using SpeX (Rayner et al. 2003), respectively. The combined optical/NIR spectrum presented by Glikman et al. provides the distinct double-peaked profiles of H$\alpha$ and H$\beta$ emission lines and the unresolved He I $\lambda10830$ emission line. For the high-quality profile of He I $\lambda10830$, we obtained an improved NIR spectrum with the Palomar TSpec spectrograph on May 25$\rm ^{th}$, 2015. Four exposures of 400 seconds each in an A-B-B-A dithering model were performed with a slit width of 1.0 arcsec. The follow-up data were calibrated using the A0V star HD145647.
The combined archive spectrum shows two broad-line emission systems. The higher redshifted `red-system' at $z=0.9880$ shows typical broad (H$\alpha$, H$\beta$, H$\gamma$ and Mg II) and narrow ([O II], [O III] and [Ne III]) lines. The lower redshifted `blue-system' at $z=0.9575$ shows broad Balmer lines. The combined archive spectrum and the high-quality H$\alpha$ emission line are shown in the top panel of Figure \ref{f1s-J1507}. Furthermore, the new NIR spectrum also clearly shows a double-peaked broad He I $\lambda10830$ emission line in the $K-$band. Similar to the emission-line analysis of SDSS J1320+5747, the emission-line profiles of SDSS J1507+3129 with the underlying continua removed are shown in the bottom panels of Figure \ref{f1s-J1507}. The emission-line profile of He I $\lambda10830$ is multiplied by 7. H$\beta$ presents almost the same profile as H$\alpha$. The same fitting process Tang and Grindlay (2009) used for the H$\alpha$ profile was also used for SDSS J1507+3129. The best-fit disk-line parameters are as follows:  $i \sim 50^\circ$, $r_1 \sim 650\ r_G$, $r_2 \sim 10,000\ r_G$, $\rm \sigma=1200\ km\ s^{-1}$, and $q=-3$. The velocity shifts of the blue-system and the broad and narrow components in the red-system are $-4610$, 0 and $\rm 0\ km\ s^{-1}$ with the $FWHM$ widths of 2005, 2241, and $\rm 589\ km\ s^{-1}$, respectively. In the left panel of Figure \ref{f1s-J1507}, the best-fit components and the sum of the H$\alpha$ emission line are shown in cyan and red. Moreover, we also plot the disk-line component and the modeled profile of H$\alpha$ in the He I $\lambda10830$ panel.

\figurenum{A1}

\figurenum{A2}

\end{document}